\pdfoutput=1

\documentclass[11pt]{article}

\usepackage{acl}

\usepackage{times}
\usepackage{latexsym}

\usepackage[T1]{fontenc}

\usepackage[utf8]{inputenc}

\usepackage{microtype}

\usepackage{inconsolata}

\usepackage{graphicx}

\title{Process-Supervised Reinforcement Learning for Code Generation}

\author{
 \textbf{Yufan Ye\textsuperscript{1}},
 \textbf{Ting Zhang\textsuperscript{2}},
 \textbf{Wenbin Jiang\textsuperscript{2}},
 \textbf{Hua Huang\textsuperscript{2,}\thanks{Corresponding author}}
\\
\\
 \textsuperscript{1}Beijing Institute of Technology,
 \textsuperscript{2}Beijing Normal University
\\
  \href{}{yeyufan@bit.edu.cn},\href{}{
  \{tingzhang, jiangwenbin, huahuang\}@bnu.edu.cn}
}

\usepackage{amsmath}
\usepackage{array}
\usepackage{booktabs}
\usepackage{graphicx} 
\usepackage{adjustbox} 
\usepackage{subfigure}
\usepackage{multirow}
\usepackage{bbm}
\usepackage{CJKutf8}
\usepackage{microtype}
\usepackage{xcolor}   
\definecolor{darkgreen}{RGB}{0,100,0}
\definecolor{crimson}{RGB}{220,20,60}
\usepackage{algorithmic}
\usepackage{algorithm}

\begin{document}
\maketitle
\begin{abstract}
Existing reinforcement learning strategies based on outcome supervision have proven effective in enhancing the performance of large language models(LLMs) for code generation. 
While reinforcement learning based on process supervision has shown great promise in handling multi-step reasoning tasks, its effectiveness in code generation remains largely underexplored and underjustified.
The primary obstacle stems from the resource-intensive nature of constructing high-quality process-supervised data, which demands substantial human expertise and computational resources.
In response to this challenge, we propose a "statement mutation/refactoring-compile and execution verification" strategy: mutating and refactoring code line-by-line through a teacher model, and utilizing compiler execution results to automatically label each line, resulting in line-by-line process-supervised data, which is pivotal for training a process-supervised reward model. The trained reward model is then integrated into the PRLCoder framework, followed by experimental validation on several benchmarks. Experimental results demonstrate that process-supervised reinforcement learning significantly surpasses methods relying solely on outcome supervision. Notably, in tackling complex code generation tasks, process-supervised reinforcement learning shows a clear advantage, ensuring both the integrity of the code generation process and the correctness of the generation results.

\end{abstract}

\begin{figure*}[t]
  \centering
  \includegraphics[width=1.0\textwidth]{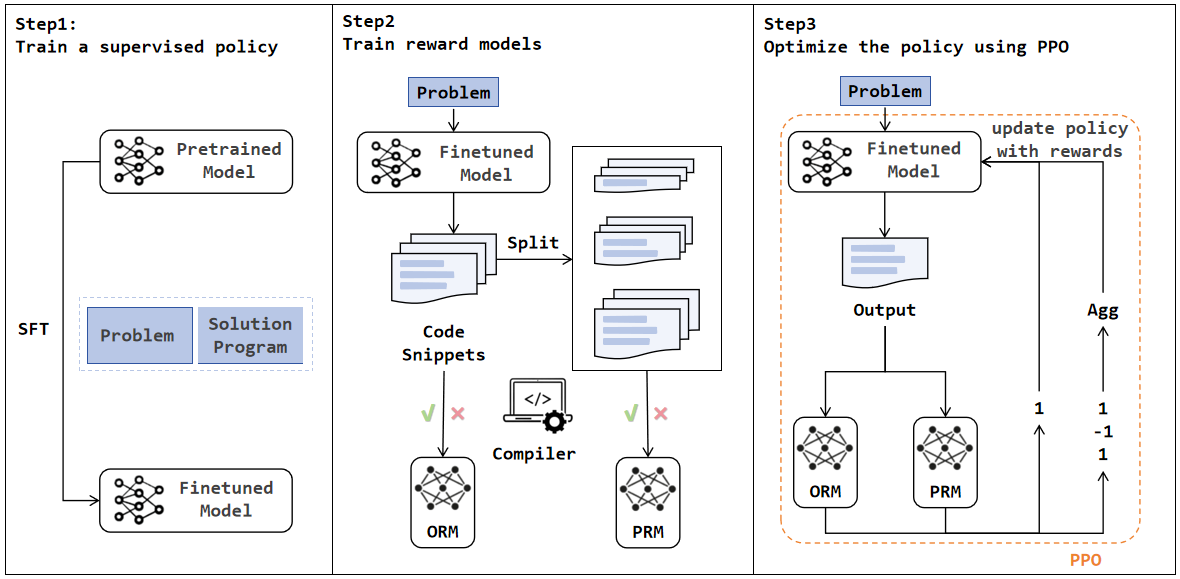}
  \caption{Illustrating the overall framework of our PRLCoder with three-phase structure: supervised training, reward model training (including ORM for comparison), and reinforcement learning employing the trained reward model.}
  \label{fig:overall process}
\end{figure*}

\section{Introduction}
Automatic code generation refers to the process of writing code automatically through algorithms or programs. Traditionally, automatic code generation has relied primarily on rule-driven programming tools and template-based code generators \citep{little2007keyword,gvero2015interactive}. These tools are typically only capable of handling simple, highly repetitive tasks and required developers to precisely define rules and logic. In recent years, with the emergence of LLMs based on deep learning and natural language processing (such as GPT \citep{brown2020language,floridi2020gpt,achiam2023gpt} and LLaMA \citep{touvron2023llama,touvron2023llama2,dubey2024llama}), the capabilities of automatic code generation have been substantially improved. These models can understand natural language descriptions and automatically generate corresponding code \citep{li2023starcoder}, even solving complex programming problems \citep{allamanis2018survey,zan2022large}, thereby greatly enhancing development productivity.

To better align models with complex human demands, reinforcement learning (RL) has played a crucial role by integrating human feedback \citep{ouyang2022training,lee2023rlaif}. The strength of RL lies in its ability to indirectly optimize non-differentiable reward signals, such as CodeBLEU scores \citep{ren2020codebleu} and human preferences \citep{wu2023human}, through policy optimization and value function approximation \citep{williams2017hybrid,dhingra2016towards}. However, obtaining the required human feedback often demands significant human effort and resources \citep{casper2023open}. In code generation tasks, reinforcement learning demonstrates unique advantages: language models can automatically utilize compiler feedback from unit tests as reward signals, reducing excessive reliance on human feedback \citep{zhang2023survey,le2022coderl,wang2022compilable,shojaee2023execution}. This approach not only efficiently optimizes the output but also significantly enhances the model's performance in code generation tasks.

Although these methods have achieved great success, they predominantly rely on compiler feedback signals from entire code segments to train the reward model, namely Outcome-Supervised Reward Model (ORM), raising the sparse reward space issue \citep{russell2016artificial,amodei2016concrete}, where the policy has no idea how well
it is performing during the training before reaching the ultimate output. In this context, Process-Supervised Reward Model (PRM) \citep{uesato2022solving,lightman2023let} offers a new perspective. This model provides step-level feedback for multi-step reasoning results generated by language models, helping to identify and correct errors in intermediate steps, rather than focusing solely on the final outcome. However, the current PRM has only been validated in the field of logical reasoning and has yet to demonstrate its effectiveness in code generation. Moreover, given the high cost of manual labeling required to construct datasets for training PRMs, efficiently building a process-supervision dataset tailored for code generation remains a critical challenge.

In this paper, we introduce PRLCoder, an innovative framework leveraging process-supervised reinforcement learning to enhance code generation, as depicted in Figure \ref{fig:overall process}, which outlines its three-phase structure: supervised training, reward model training (including ORM for comparison), and reinforcement learning employing the trained reward model.
Importantly, we design a "statement mutation/refactoring-compile and execution verification" strategy to automatically generate process-supervised data. To be specific, for each line of the code, we adopt a teacher model to perform mutation and refactoring operations, where mutations produce code with different functionality from the original statement, while refactoring aims to preserve the statement's functionality as much as possible. The modified block of code is then verified by a compiler, and based on the outcome of test cases, the samples are labeled as either "Positive" or "Negative". We observe that the test cases in the MBPP dataset exhibit low coverage and are unable to accurately label the sample code. To address this, we extend the test cases to more effectively leverage compiler feedback signals. Finally, we utilize the trained PRM to assign fine-grained rewards to each line of code, enabling reinforcement learning.
This method not only significantly reduces the time and cost required for manual annotation in traditional process supervision but also eliminates errors and biases in manual labeling. Additionally, the accuracy of fine-grained rewards makes the model more efficient in environment exploration, enhancing the stability of the training process.

We evaluate our approach on two widely used benchmark datasets, MBPP and HumanEval. Experimental results indicate that PRLCoder improved the pass rate by 10.5\% compared to the base model and by 5.1\% compared to outcome-supervised reinforcement learning, with more significant performance gains in tasks involving complex code generation. In summary, our main contributions are as follows:
\begin{itemize}
\item[1)] 
To the best of our knowledge, we present the first attempt to investigate process-supervised reinforcement learning in the realm of code generation, exploring its potential to enhance the performance.
\item[2)] 
To address the challenge of the resource-intensive manual labeling process, we introduce a "mutation/refactoring-verification" strategy to automatically generate high-quality process-supervised data for training reward models. Additionally, we supplement the test cases in the MBPP dataset to improve the accuracy of line-level labeling.
\item[3)] 
Empirically we demonstrate that process supervision outperforms outcome supervision in code generation, achieving 4.4\% improvement with more significant performance gain in tasks involving complex code generation.
\end{itemize}

\section{Related Work}
\subsection{Pretrained LLMs for Code}
As LLMs begin to exhibit early signs of artificial intelligence, their applications have extended beyond text processing. In the domain of code generation, LLMs, trained on extensive corpora of code and natural language, are capable of generating code that is coherent both syntactically and semantically \citep{jiang2024survey,guo2020graphcodebert,li2022competition,nijkamp2022codegen}. Among them, encoder models like CodeBERT \citep{feng2020codebert} focus on understanding code structure and semantic relationships, encoder-decoder models like CodeT5 \citep{wang2021codet5} specialize in translating high-level language descriptions into concrete code, while decoder-only models like DeepSeekCoder \citep{guo2024deepseek} generate syntactically correct and semantically coherent code through autoregressive methods. Additionally, researchers in the coding community have applied instructional tuning to their models. \citet{wang2023codet5+} fine-tuned CodeT5+ using 20,000 instruction data generated by InstructGPT, resulting in InstructCodeT5+ with enhanced generalization capabilities. However, these models largely overlook the unique sequential features of code, exhibiting limited performance in handling complex issues and in cross-task generalization and scalability \citep{zhang2024deep}.

\subsection{RL based on Compiler}
Reinforcement learning is a method of learning through "trial and error," aiming to enable an agent to interact with the environment and receive rewards to guide behavior and maximize cumulative rewards \citep{mnih2013playing,mnih2015human,van2016deep}. Given the requirement for both syntactic and functional correctness in code generation tasks, leveraging compiler feedback signals from unit tests for reinforcement learning has become a more competitive strategy. CodeRL \citep{le2022coderl} takes advantage of this by introducing a critic network to predict the functional correctness of generated programs, providing dense feedback signals to the code generation model (i.e., the actor network) for reinforcement learning. Similarly, CompCoder \citep{wang2022compilable} and PPOCoder \citep{shojaee2023execution} employ the Proximal Policy Optimization (PPO) algorithm to train CodeGPT and CodeT5, respectively, while RLTF \citep{liu2023rltf} uses compiler-generated error messages and locations to provide more fine-grained feedback. It constructs an online reinforcement learning framework with multi-granularity unit test feedback, generating data in real-time during the training process. However, despite the progress made by these outcome-supervised reinforcement learning methods, they still face challenges such as sparse reward space and training instability.

\subsection{Process Supervision}
Outcome supervision focuses on the final output, whereas process supervision provides guidance through intermediate steps \citep{uesato2022solving,luo2024improve,wang2024math}. 
 \citet{lightman2023let} collected a large amount of process-supervised data and constructed the PRM800K dataset. The results demonstrated that process supervision significantly outperformed outcome supervision in solving problems in the MATH dataset. \citet{wu2024fine} conducted further experiments using fine-grained human feedback as explicit training signals for tasks such as detoxification and long-form question answering. Their study showed that fine-grained feedback provides more effective supervision signals compared to holistic feedback on long texts. In the coding domain, \citet{ma2023let} modified atomic operators by employing AST to train a reward model, which was applied in multi-step reasoning and proven effective. \citet{dai2024process} utilized LLMs to generate completions for code prefixes and employed automated testing to evaluate their correctness. Based on this evaluation, they determined whether the prefixes were correct and then automatically generated a process-supervised dataset, exploring the effectiveness of process supervision in the code domain. Compared with the work we conducted during the same period, there are differences in the core aspect of automatically creating the process-supervised dataset. Since they used a closed-source model, we cannot directly compare the advantages and disadvantages of the two methods. However, it is still particularly important to explore more optimized mutation/refactoring mechanisms, train more reliable PRM, and further study the potential advantages of reinforcement learning based on process supervision over outcome supervision in the coding domain. 

\begin{figure*}[t]
  \centering
  \includegraphics[width=1.0\textwidth]{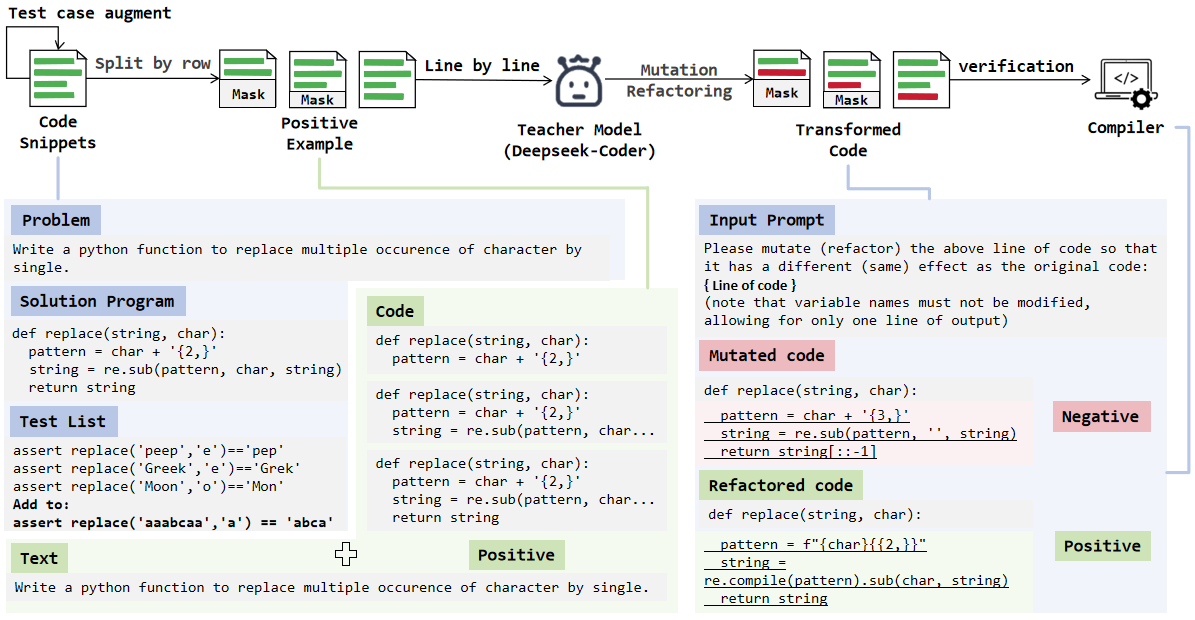}
  \caption{The schematic diagram of the method for automatically constructing the reward dataset for process supervision in the field of code generation. The bolded portions represent code statements that have been mutated or refactored by DeepSeek-Coder-V2, and the subsequent statements will undergo mask processing. 
  }
  \label{fig:dataset construction}
\end{figure*}

\section{Approach}
In this section, we will elaborate on the methodological details of PRLCoder. By offering more fine-grained rewards, PRLCoder enables the PPO reinforcement learning algorithm to explore and optimize more accurately in code generation. 

\subsection{Process-Supervised Dataset Construction}
Similar to the field of mathematical logic reasoning, collecting fine-grained human feedback through manual annotation to construct step-level reward datasets often requires significant human and material resources. To address this, we propose an innovative approach that leverages a teacher model and compiler feedback to automatically construct a process-supervised reward dataset for the domain of code generation. Figure \ref{fig:dataset construction} illustrates a schematic of the dataset generation process.


Formally, let $\mathcal{D} = \{p_i,s_i\}_{i=1}^N$
denotes the code generation training dataset, where 
$p_i$ represents the $i$-th problem description and 
$s_i$ is the corresponding solution program code snippet. Initially,
we leverage this reference code to construct positive samples. To be specific, we segment the reference code line by line, resulting in $s_i=\{s_{i1},\cdots,s_{iL_i}\}$ with $L_i$ being the number of lines. Then for each line of code, all subsequent lines are masked, and we directly mark the corresponding label for the line as \textbf{"positive"}. In other words, the original reference code can be directly reformulated as positive samples for process supervision with the format: $\{(p_i,s_{ij|j \leq l}),\textbf{"positive"}; l=1,\cdots, L_i\}_{i=1}^N$.

Positive samples alone are insufficient for training reward models; hence, we design a novel strategy to construct negative samples. Specifically for each line of code,
we employ a teacher model to perform mutate and refactoring operations using specific prompt examples detailed in Figure \ref{fig:dataset construction}. The modified line, along with the remaining code, is then validated through the compiler. Based on the compiler feedback, it is labeled as \textbf{"positive"} if it passes all test cases, or \textbf{"negative"} otherwise.


It is worth noting that during this process, we discover
that in the MBPP dataset, the modified line can still pass all the test cases despite containing errors for certain problem descriptions. This issue arises due to insufficient test case coverage. In order to construct a more precise step-level reward dataset, we expand these test cases. More details about test case extension can be found in Section \ref{sec:MBPP}. Subsequently, we follow the PPO reinforcement learning algorithm to optimize both the policy model and the value model.

\subsection{Reward Model Training}
\textbf{Outcome-Supervised Reward Model.} ORM adopts a holistic reward approach, mapping the overall quality and reliability metrics corresponding to the problem description $d$ and the generated code $w$ into a single scalar reward. Typically, this reward is only assigned to the final token in the generated sequence and is defined as follows:

\begin{equation}
r_t^O = \begin{cases}
R_O(d, w; \theta), & t=T \\
0, & \text{otherwise} \\
\end{cases}
\end{equation}

where $\theta$ represents the parameters of ORM $R_O$. We first use the dataset constructed in the previous section to train an original ORM. However, relying solely on this dataset to train the ORM has limitations: the active learning strategy exhibits a strong bias towards incorrect answers in the dataset, thereby diminishing the overall performance of the model. Thus, we aim to explore alternative approaches to build a more robust ORM baseline.

\begin{figure}[t]
  \centering
  \includegraphics[width=0.5\textwidth]{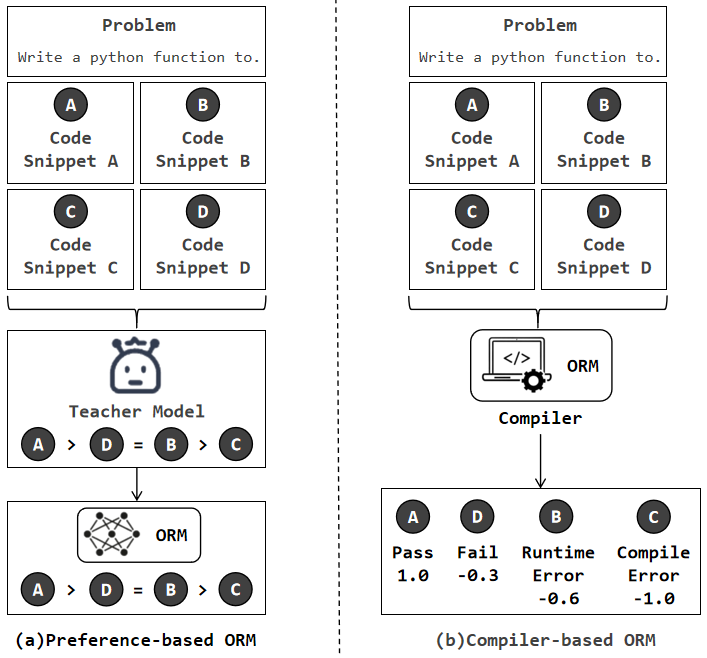}
  \caption{Training of two types of ORM.}
  \label{fig:orm}
\end{figure}

Inspired by RLHF, we design a preference-based ORM. Specifically, for each question, we uniformly sample multiple code snippets from the generator and use a teacher model to simulate human annotators ranking them based on code quality, thereby training the reward model. Moreover, to more comprehensively evaluate the advantages and disadvantages of process supervision and outcome supervision in the coding domain, we refer to methods such as CodeRL mentioned earlier. We introduce the compiler as a source of supervision signals and use four types of feedback signals generated by the compiler to optimize the generator model, thus constructing a compiler-based ORM. Figure \ref{fig:orm} illustrates the structures of these two ORM models, respectively.

\textbf{Process-Supervised Reward Model.} Our PRM rewards the quality of each code segment, allowing for finer adjustments and feedback at each step. We divide the code sequence $w$ into $k$ segments $(w_1,w_2,...,w_k)$, where $w_i$ represents the preceding part of the code sequence. The synchronous execution concludes at time $T_i$, denoted as $a_{T_i}$ = `\textbackslash n'. Within this framework, the reward model assigns a reward to each input segment $(d, w_i)$, distributing the highest reward to the final segment of $w$. Finally, the reward $r_t$ is defined as:

\begin{equation}
r_t^P = \sum_{i=1}^{k} R_P(d, w_i; \phi) \cdot \mathbbm{1}(t = T_i)
\end{equation}

where $\phi$ represents the parameters of PRM $R_P$. We use the process-supervised dataset constructed with the "mutation/refactoring-verification" strategy to train our PRM. Under this setting, the PRM's training data does not intersect with compiler-based ORM, making it difficult to compare results directly. Nevertheless, the PRM and ORM both represent the optimal results under their respective training methods.

\subsection{Reinforcement Learning Algorithm}
PPO (Proximal Policy Optimization) is a reinforcement learning algorithm based on policy gradients. Its core idea is to limit the magnitude of changes between the old and new policies to prevent excessively rapid updates \citep{schulman2017proximal,huang2024ppo}. This is particularly crucial in the code generation process, as the stability of the generated results directly impacts the quality and consistency of the code.

In code generation tasks, the PPO algorithm first interacts with the environment using the current policy $\pi_{\theta_k}$ to obtain the state $s_t$, selects an action $a_t$, and receives a reward $r_t$ and other data. Subsequently, the advantage function $A_t = \sum_{t' > t} \gamma^{t' - t} (r_{t'} + \gamma V_{\psi}(s_{t'+1}) - V_{\psi}(s_{t'}))$ is calculated for each time step, where the value function $V_{\psi}(s_t)$ represents the expected cumulative rewards from state $s_t$. See Appendix for more details. In addition, we adopt the method from \citep{wu2021recursively} to add a divergence penalty to each token, representing the ratio of the old and new policies. our reward function becomes:

\begin{equation}
r_t = r_t^P - \beta \log \frac{\pi_{\theta}(a_t | s_t)}{\pi_{\theta_{old}}(a_t | s_t)}
\end{equation}

See Appendix \ref{app:algorithm} for more details.

\section{Experiments}
\subsection{Benchmarks}
\label{sec:MBPP}

\textbf{MBPP.} To train our PRM, we first select MBPP \citep{austin2021program} as the seed dataset. The MBPP dataset consists of 974 crowdsourced Python programming problems. Each problem includes a task description, a code solution, and three automated test cases. We adopt the same prompt format as \citet{austin2021program} to prepare the input sequence: problem description + "Your code should satisfy these tests:" + three assertion.

To maximize path coverage and improve the quality of process-supervised data, we leverage LLMs to supplement the test cases in the dataset. The resulting augmented dataset is referred to as MBPP\textsuperscript{+}. See Appendix \ref{app:dataset} for more details.

\noindent
\textbf{HumanEval.} To further evaluate the framework we proposed, we also employ an additional Python program synthesis dataset of comparable difficulty. The HumanEval dataset consists of 164 original programming problems, with some problems being comparable in difficulty to fundamental software interview questions. To verify the model's generalization ability, we conduct a comprehensive evaluation of the model on the HumanEval benchmark test set.

\subsection{Settings}

\noindent
\textbf{Evalution Metric.}
We follow the method proposed by \citet{kulal2019spoc, chen2021evaluating} to evaluate function correctness using the pass@k metric, which involves generating k code samples for each problem. If any of the code samples pass the unit tests, the problem is considered correctly solved, and we report the overall proportion of problems solved. For each task, we generate n $\geq$ k samples (in this paper, we used n = 200) and calculate the number of correct samples that passed the unit tests, denoted as c $\leq$ n. The formula used in the paper is: $\text{pass@k} := \mathbbm{E}_{\text{Problems}} \left[ 1 - \frac{\binom{n-c}{k}}{\binom{n}{k}} \right]$.

\noindent
\textbf{Implementation Details.}
We fine-tune the CodeT5+ \citep{wang2023codet5+} as the policy model. During the supervised fine-tuning (SFT) phase, due to the small size of the MBPP training set, we employ a learning rate of 2e-5 and train the model for 60 epochs, taking approximately 60 minutes on a single NVIDIA A800 80G GPU. We select Unixcoder \citep{guo2022unixcoder} as the base model for ORM and PRM, training it for 10 epochs with  weight\_decay set to 0.01 and warmup\_ratio set to 0.01. In the final PPO training phase, the value model is based on T5\_base \citep{raffel2020exploring}. For each sample, we generate four code snippets using nucleus sampling with a temperature of 1.2, top-p set to 0.95, and a maximum output token count of 512. During the decoding phase, the sampling temperature for MBPP is set to 1.2, while for HumanEval it was set to 0.6, 0.8, and 1.2.

\noindent
\textbf{Training Data.}
We re-partition the dataset to improve the generalization capability and robustness of the process-supervised reward model. Specifically, we select IDs in the range 601–974 from MBPP as the training set for the SFT phase and the seed set for the process-supervised reward dataset, IDs in the range 101–500 as the training set for the RL phase, and IDs in the range 501–600 and 1–100 as the validation and test sets, respectively. The process-supervised reward dataset, generated through the automated "mutation/refactoring-verification" strategy, includes training, validation, and test subsets. The positive and negative samples in each subset are distributed as follows: 3,469/2,674 for the training set, 632/507 for the validation set, and 631/488 for the testing set.

\subsection{Experimental Results}

\subsubsection{Results on MBPP}



\begin{table*}[t]
\centering
\small
\renewcommand{\arraystretch}{1.5}
\begin{adjustbox}{width=\textwidth}
\begin{tabular}{ccccccccccccccc}
\hline

\multicolumn{2}{c}{\multirow{2}{*}{\textbf{Model}}} & \multirow{2}{*}{\textbf{Size}} & \multicolumn{4}{c}{\textbf{pass@1}} & \multicolumn{4}{c}{\textbf{pass@10}} & \multicolumn{4}{c}{\textbf{pass@80}} \\

\multicolumn{2}{c}{} & & EZY & MED & HRD & all & EZY & MED & HRD & all & EZY & MED & HRD & all \\ 
\hline
\multicolumn{2}{c}{GPT}  & 8B & - & - & - & - & - & - & - & - & - & - & - & 40.6 \\

\multicolumn{2}{c}{GPT}  & 68B & - & - & - & - & - & - & - & - & - & - & - & 53.6 \\

\multicolumn{2}{c}{GPT}  & 137B & - & - & - & - & - & - & - & - & - & - & - & 61.4 \\

\multicolumn{2}{c}{CodeGen}  & 6.1B & 25.4 & 11.8 & 3.2 & 15.9 & 48.3 & 34.2 & 25.1 & 36.5 & 58.4 & 44.3 & 37.7 & 49.8 \\ 

\multicolumn{2}{c}{LLaMA}  & 7B & 27.9 & 12.6 & 4.1 & 17.7 & 50.6 & 36.9 & 26.7 & 40.3 & 68.4 & 54.8 & 48.9 & 59.3 \\
\hline

\multicolumn{2}{c}{CodeT5+} & 770M & 27.6 & 13.4 & 4.4 & 18.0 & 52.2 & 37.0 & 28.0 & 41.6 & 67.9 & 56.4 & 50.0 & 60.3 \\

\multicolumn{2}{c}{O-ORM} & 770M & \textbf{29.0} & \textbf{15.5} & \textbf{6.6} & \textbf{20.1} & 49.2 & 34.4 & 31.2 & 39.7 & 65.7 & 52.5 & 51.2 & 57.9 \\

\multicolumn{2}{c}{P-ORM} & 770M & 28.5 & 13.9 & 5.0 & 18.2 & \textbf{53.3} & 37.6 & 28.4 & 41.9 & 67.4 & 56.9 & 52.8 & 62.0 \\ 

\multirow{2}{*}{C-ORM} & CodeRL & 770M & 28.5 & 13.7 & 4.6 & 18.1 & 52.3 & 37.9 & 29.9 & 42.0 & 66.1 & 56.9 & 52.6 & 61.0 \\

& PPOCoder & 770M & 28.9 & 14.0 & 4.5 & 18.5 & 52.7 & 38.0 & 29.6 & 42.4 & 66.7 & 57.3 & 53.4 & 61.9 \\
\hline

\multicolumn{2}{c}{RSFT} & 770M & 28.4 & 13.8 & 4.9 & 18.3 & 52.5 & 37.6 & 30.4 & 42.6 & 68.4 & 58.4 & 57.2 & 62.2 \\

\multicolumn{2}{c}{PRLCoder} & 770M & 27.8 & 14.5 & 5.7 & 18.7 & 53.0 & \textbf{38.4} & \textbf{32.3} & \textbf{43.0} & \textbf{69.0} & \textbf{60.0} & \textbf{59.6} & \textbf{63.8} \\
\hline

\end{tabular}
\end{adjustbox}
\caption{Performance results for various models on MBPP\textsuperscript{+} testing set. O-ORM represents the original ORM, P-ORM represents the preference-based ORM, C-ORM represents the compiler-based ORM, and RSFT represents rejection sampling fine-tuning.}
\label{tab:MBPP}
\end{table*}

\begin{table}[t]
\centering
\small
\renewcommand{\arraystretch}{1.5}
\begin{tabular}{ccccc}
\hline
\multicolumn{2}{c}{\textbf{Model}} & \textbf{Size} & \textbf{Pass@1} & \textbf{Pass@100} \\ \hline
\multicolumn{2}{c}{GPT-J}  & 6B & 11.6 & 27.7 \\ 
\multicolumn{2}{c}{CodeGen}  & 6.1B & 10.4 & 29.8 \\ 
\multicolumn{2}{c}{LLaMA}  & 7B & 10.5 & 36.5 \\  
\multicolumn{2}{c}{CodeT5+}  & 770M & 12.5 & 38.0 \\
\multicolumn{2}{c}{O-ORM}  & 770M & 13.2 & 40.1 \\
\multicolumn{2}{c}{P-ORM}  & 770M & 12.9 & 40.6 \\  
\multirow{2}{*}{C-ORM} & CodeRL & 770M & 12.6 & 39.7 \\
& PPOCoder & 770M & 13.0 & 40.0 \\ \hline
\multicolumn{2}{c}{PRLCoder}  & 770M & \textbf{13.6} & \textbf{41.8} \\ \hline
\end{tabular}
\caption{Quantitative results on Humaneval benchmark.}
\label{tab:Humaneval}
\end{table}

\begin{figure*}[t]
  \centering
  \label{fig:reward model analysis}
  \subfigure[Accuracy, F1, Precision, Recall]{
  \label{fig:rmfig_a}
  \includegraphics[width=0.45\linewidth]{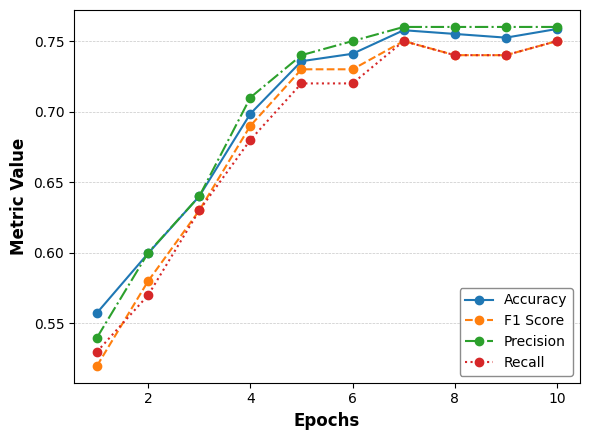}
  }
  \hspace{0.03\linewidth} 
  \subfigure[Class Accuracy]{
  \label{fig:rmfig_b}
  \includegraphics[width=0.45\linewidth]{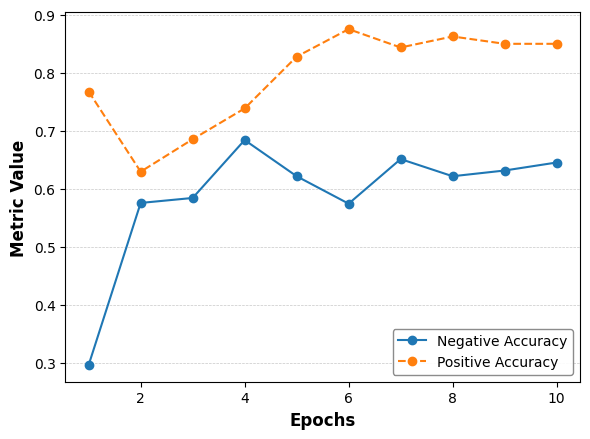}
  }
  \caption {Quantitative analysis of the process-supervised reward model for code trained using our method.}
\end{figure*}

\begin{figure}[t]
  \centering
  \includegraphics[width=0.9\linewidth]{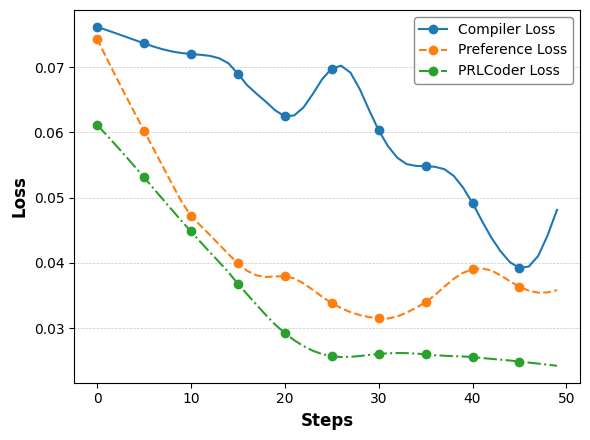}
  \caption{The loss curves of the reinforcement learning under three different supervision methods.}
  \label{fig:loss}
\end{figure}


To evaluate the performance of our PRLCoder in code generation, we conduct comprehensive experiments on the MBPP\textsuperscript{+} test set and show the experimental results in Table~\ref{tab:MBPP}. We hypothesize that process supervision may have a more significant advantage in complex code generation tasks. Therefore, to provide a more comprehensive evaluation of our PRLCoder method, we divide the MBPP\textsuperscript{+} test set into three categories based on the length of the standard\_code: <50, 50-100, and >100. This categorization roughly reflects the difficulty levels of the programming tasks, which are denoted as EZY, MED, HRD respectively in the table.

\noindent 
\textbf{Comparison with LLMs. }
For the baseline, we use GPT models with parameter sizes ranging from 4B to 137B as reported by \citet{austin2021program}, and the results are obtained from the original paper. Additionally, we evaluate the CodeGen and LLaMA models under the same experimental conditions to ensure fairness and consistency in comparison. 
It can be seen that our model achieves noticeable performance improvement with significantly smaller model size.

\noindent 
\textbf{Comparison with ORMs.}
We then evaluate our approach compared to methods based on outcome-supervised reward models (ORMs). We adopt the original ORM, the preference-based ORM and the compiler-based ORM for comprehensive evluation. 
To ensure fair comparison, we reproduce the experimental results on the MBPP\textsuperscript{+} dataset. 
Specifically for compiler-based method, we compare with two competitive methods: CodeRL and PPOCoder that utilized different reinforcement learning techniques. 
Our process supervision approach achieves significant improvements across tasks of varying difficulty levels. The improvement is particularly notable in medium and difficult problems. This indicates that process supervision can provide more detailed guidance on the model’s rewards in complex tasks, leading to the generation of more accurate code snippets.

Although the original ORM achieves higher results in Pass@1, it shows a decline in performance compared to the base model in both Pass@10 and Pass@100. We hypothesize that this drop may be due to the method's strong bias toward erroneous code in the dataset, causing it to misjudge some correct code snippets during the RL process, thereby reducing the number of correct answers. This observation suggests that relying solely on outcome-based strategies may be insufficient for improving the model's code generation capabilities comprehensively. In contrast, process supervision, by guiding the intermediate steps in the generation process, can mitigate this issue.

\noindent
\textbf{Application of PRM to Rejection Sampling.}
To further validate the effectiveness of PRM, we design and conduct rejection sampling fine-tuning experiments independent of any RL, thereby evaluating the applicability and performance of PRM beyond the constraints of reinforcement learning.

\subsubsection{Results on HumanEval}

To further assess the generalization capability of PRLCoder, we also test the performance of the MBPP\textsuperscript{+} fine-tuned model on the HumanEval dataset. The specific results are shown in Table \ref{tab:Humaneval}. The results indicate that the PRLCoder outperforms outcome-supervised approaches. It is worthmentioning that our CodeT5+ performing below the original results reported on the HumanEval dataset is due to our fine-tuning being conducted solely on the MBPP\textsuperscript{+} training set, which utilizes significantly less data than the dataset used in the original study.

\subsection{Analysis}
We conduct an analysis of the automatically constructed step-level dataset, focusing on evaluating its performance in training the PRM, as well as the model's efficiency and stability in RL training.


\noindent
\textbf{The Classification Accuracy of Reward model.} Based on the constructed dataset, we train PRM, as shown in Figure \ref{fig:rmfig_a}. During the training phase, the overall accuracy of the model reaches nearly 80\%. To further evaluate the model's performance, we analyze the classification accuracy for "Positive" and "Negative" labels during the training process, with the results presented in Figure \ref{fig:rmfig_b}. The model achieves an overall performance of 0.76 on the test set, with specific classification results of ([0.84, 0.65]). These findings indicate that the model demonstrates strong performance in the reward-based code generation task, validating the effectiveness of the proposed training strategy and providing robust support for further optimization of the code generation process. Additionally, in Appendix \ref{app:case}, we present a case study where the model evaluates the code generation results with line-by-line rewards, further demonstrating the model's effectiveness in optimizing the code generation process.

\noindent
\textbf{Training process.} When using the PPO reinforcement learning algorithm for model training, we compare the train\_loss curves under three different supervision methods, as shown in Figure \ref{fig:loss}. The results indicate that our proposed PRLCoder method demonstrates faster convergence during training and exhibits significantly higher stability compared to the other two outcome supervision methods. This suggests that process supervision not only improves training efficiency in code generation models but also significantly enhances the stability of the training process.

\section{Conclusion}
In this paper, we propose a novel approach called PRLCoder, which presents the first attempt to enhance the code generation by process reward models, which provide intermediate reward signals. To tackle the challenge of costly labeling, we design an innovative step-level dataset construction strategy that automatically generates dataset for training the code PRM using feedback from a teacher model and a compiler. We also discover the low coverage of current test cases in MBPP and perform augmentation to enable effective PRM training. 
Experimental results show that on the MBPP and HumanEval datasets, our method significantly improves the quality of code generation. Our approach successfully validate the superiority of PRMs over ORMs in code generation, most notably without the need for resource-intensive manual labeling.

\section{Limitations}
Looking ahead, several aspects of PRLCoder can be further optimized and expanded. First, the current seed dataset has limited diversity, which may hinder the generalization capability of the trained PRM. Future research could consider utilizing more rich and diverse seed datasets to better cover various scenarios and requirements in code generation. Additionally, current experiments with PRLCoder have only been conducted on CodeT5+, and future work could explore its applicability and performance across more types and larger-scale code generation models. Furthermore, our proposed "mutation/refactoring-verification" strategy is not only applicable to code generation but also has the potential to establish process-supervised mechanisms for other reasoning or planning tasks. Future studies could further investigate the applicability and advantages of this strategy in other fields, especially its potential in addressing complex reasoning and planning challenges.

\bibliography{custom}
\clearpage

\appendix



\section{PPO Algorithm}
\label{app:algorithm}
The full algorithm of PRLCoder is detailed in Algorithm \ref{alg:PPO}.

\section{Dataset Augmentation}
\label{app:dataset}
\begin{figure*}[t]
  \centering
  \includegraphics[width=0.9\textwidth]{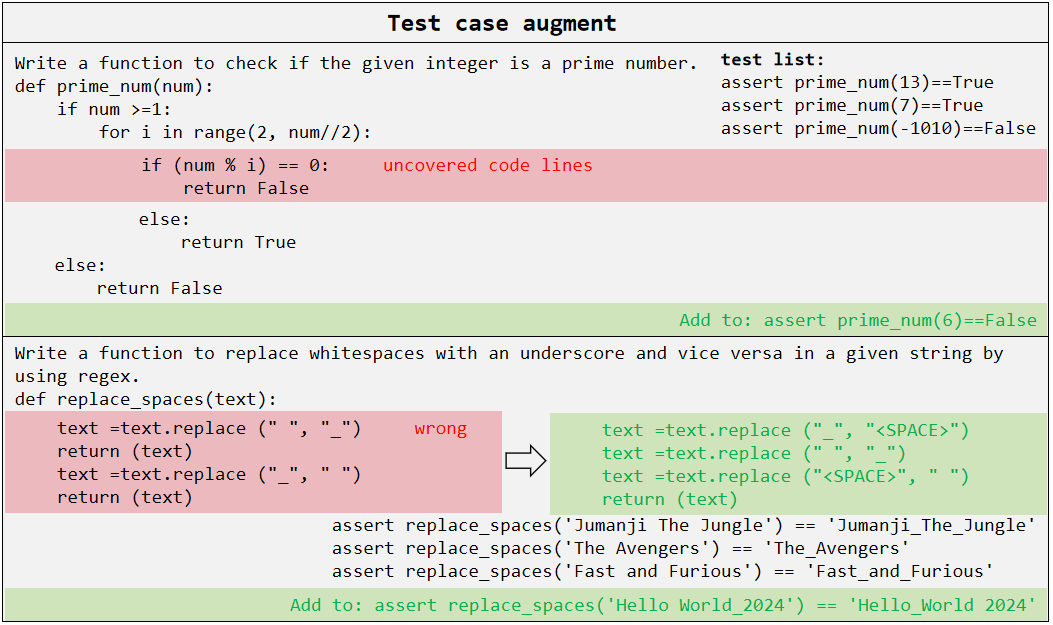}
  \caption{Some examples of the modifications we made to MBPP to align with our method}
  \label{fig:MBPP+}
\end{figure*}

To establish a more standardized PRM dataset, we first normalize the code by standardizing the use of `\textbackslash t', ensuring uniform code formatting. Subsequently, to fully leverage feedback signals provided by the compiler, we supplement test cases for MBPP problems. Specifically, we leverage a teacher model to generate test cases aimed at achieving comprehensive path coverage. These test cases are executed using a compiler to verify their correctness and effectiveness. The prompt provided to the teacher model is designed as follows: "Given the following code and its existing test cases, supplement with a new test case to achieve full path coverage." As shown in Figure \ref{fig:MBPP+}, here are some examples of the modifications we made to MBPP.

\section{Case Study}
\label{app:case}

\begin{figure*}[t]
  \centering
  \includegraphics[width=0.9\textwidth]{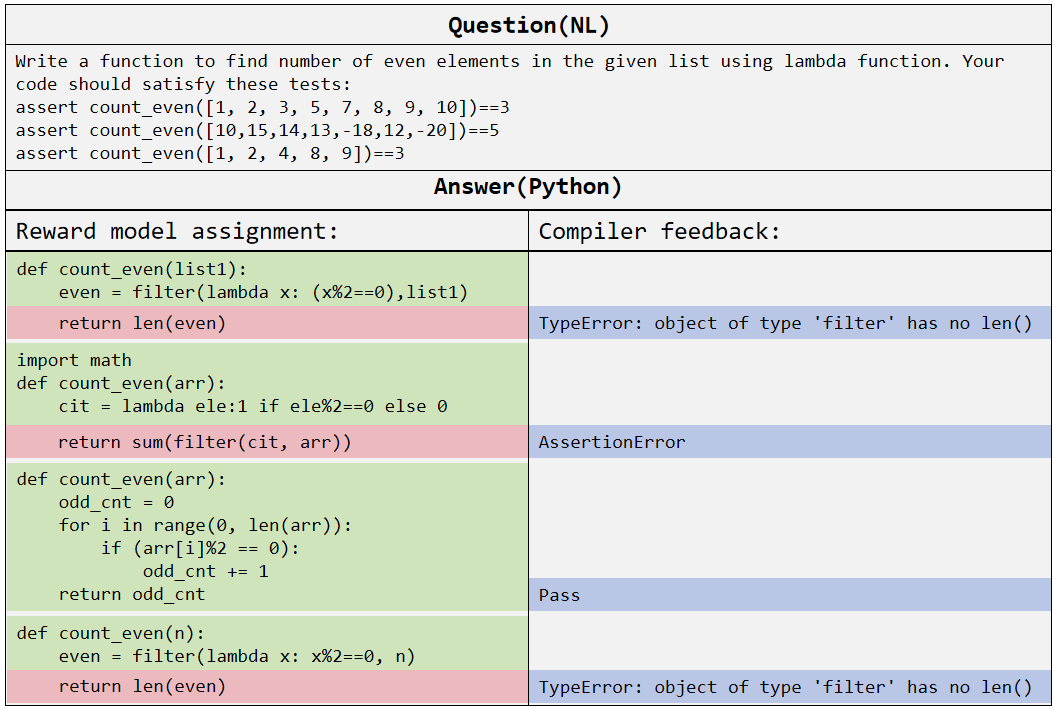}
  \par\vspace{3cm}
  \includegraphics[width=0.9\textwidth]{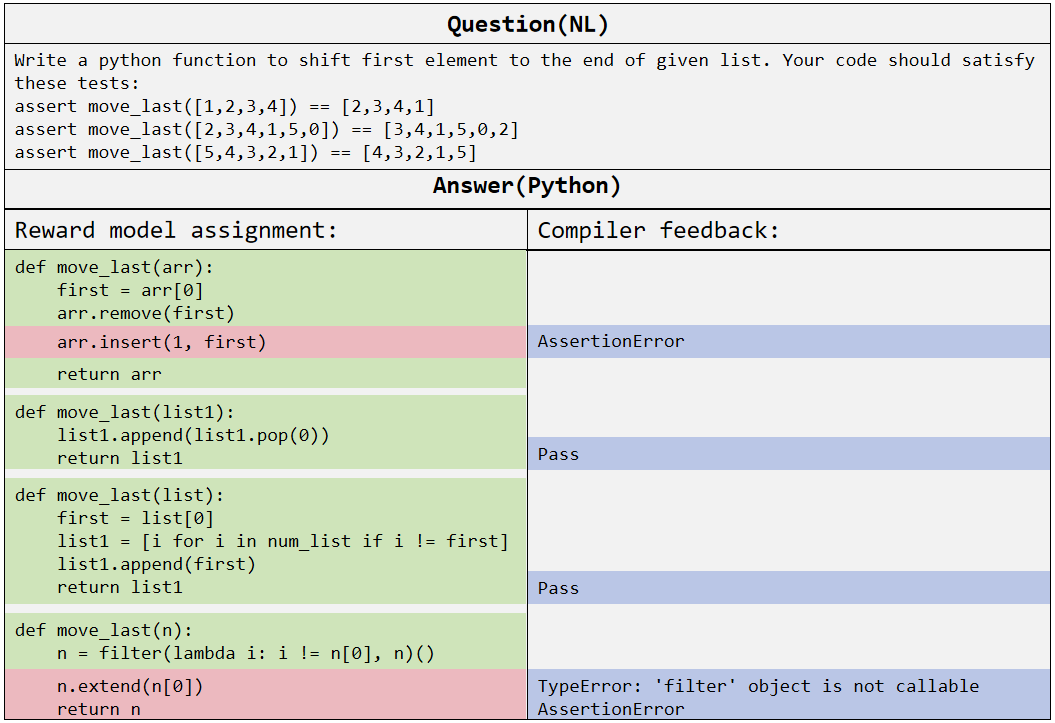}
  \caption{case study on assigning rewards line by line in our PRM}
  \label{fig:reward}
\end{figure*}

\textbf{Reward model.} We conduct line-by-line reward evaluation experiments on the code generation results using the trained process-supervised reward model (as shown in the Figure \ref{fig:reward}). The experimental results show that the model's reward evaluations are largely consistent with the feedback from the compiler, accurately assigning negative rewards to erroneous lines of code. This effectively enhances the model's ability to assess the quality of code generation.

\begin{figure*}[t]
  \centering
  \includegraphics[width=0.9\textwidth]{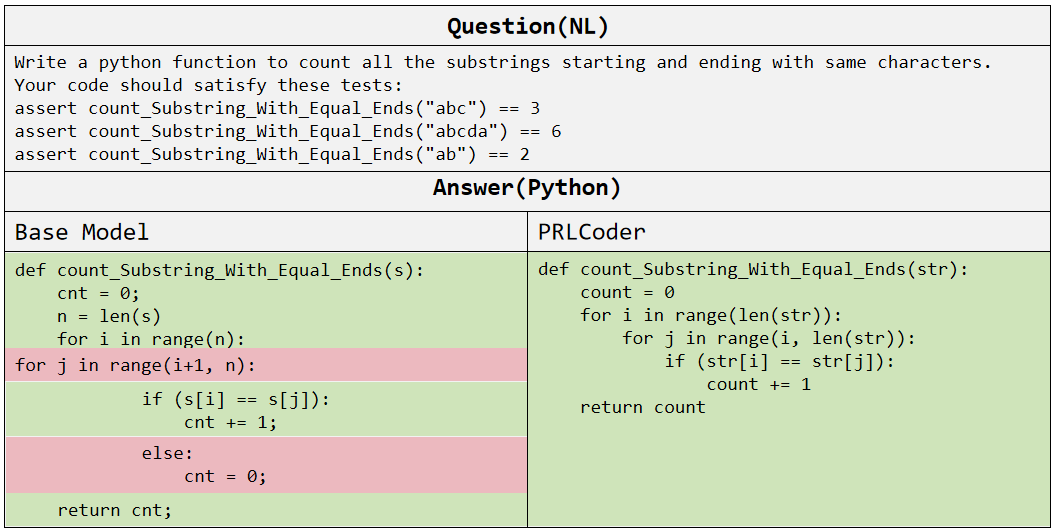}
  \par\vspace{3cm}
  \includegraphics[width=0.9\textwidth]{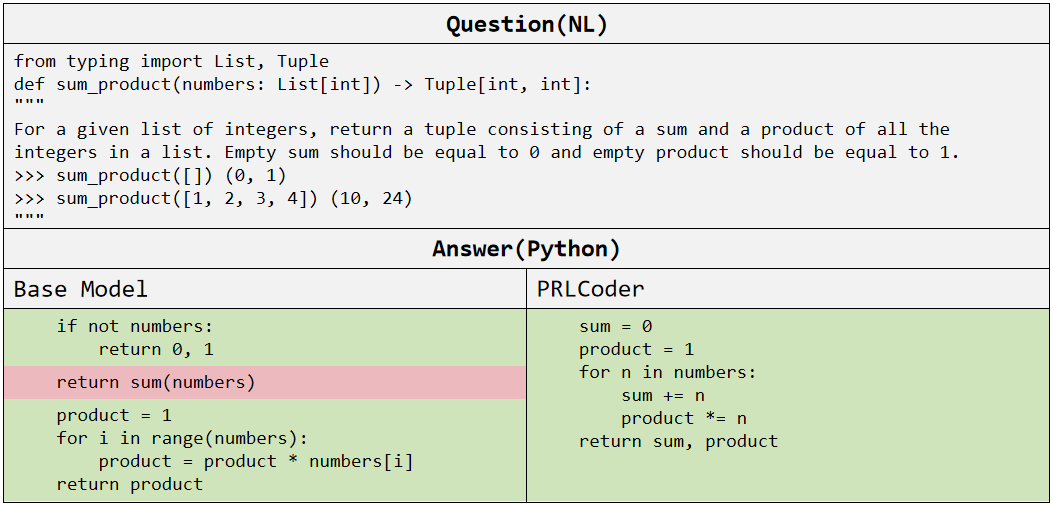}
  \caption{case study on code generation results of the base model and PRLCoder}
  \label{fig:policy}
\end{figure*}

\textbf{Policy model.} We conduct a comparative analysis of the code generated by the baseline model and PRLCoder. As shown in Figure \ref{fig:policy}, the code generated by PRLCoder not only maintains process integrity but also ensures greater accuracy of the results. This demonstrates that PRLCoder's modeling of process supervision in code generation tasks is more effective, thereby enhancing the quality and reliability of the generated code.

\section{Error Distribution}
\label{app:distribution}

\begin{figure*}[t]
  \centering
  \includegraphics[width=0.9\textwidth]{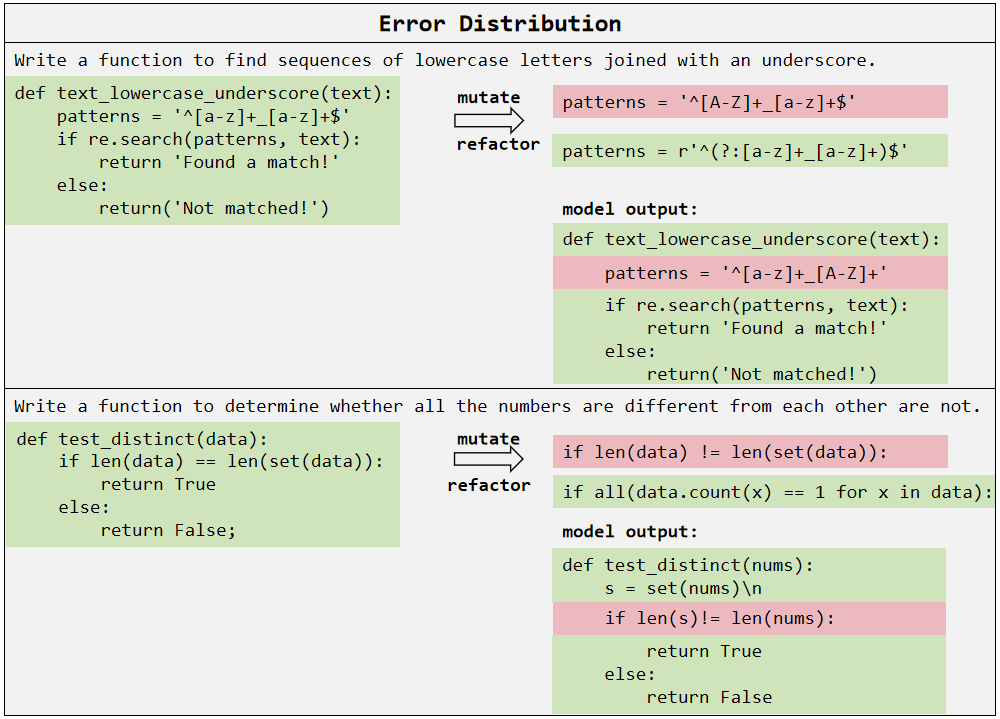}
  \caption{Some examples of the same error distribution generated by the reward dataset and the base model.}
  \label{fig:error}
\end{figure*}

To validate the effectiveness of our proposed strategy, we conduct an error distribution analysis on the automatically constructed reward dataset and the code generated by the baseline model. As shown in Figure \ref{fig:error}, the error distributions of the two code sets exhibit significant overlap, demonstrating that the reward dataset constructed using this strategy effectively captures common error patterns in the code generation process. Furthermore, when this dataset is used to train PRM within a reinforcement learning framework, it significantly enhances the model's ability to supervise code generation.


\begin{algorithm*}[t]
    \caption{Process-Supervised Reinforcement Learning for Code Generation}
    \label{alg:PPO}
    \renewcommand{\algorithmicrequire}{\textbf{Input:}}
    \renewcommand{\algorithmicensure}{\textbf{Output:}}
    \begin{algorithmic}[1]
        \REQUIRE initial policy model $P_{\theta_\text{init}}$; initial value model $V_{\psi_\text{init}}$; PRM $R_{\phi}$ trained from step-level datasets; code task prompts $\mathcal{D}$; hyperparameters $\gamma, \lambda, \epsilon, \beta$
        \ENSURE $P_{\theta}$
        
        \STATE policy model $P_\theta \gets P_{\theta_\text{init}}$, value model $V_\psi \gets V_{\psi_\text{init}}$
        \FOR{step $= 1, \dots, M$}
            \STATE Sample a batch $\mathcal{D}_b$ from $\mathcal{D}$
            \STATE Sample output sequence of program $w^n \sim P_\theta(\cdot \mid x^n)$ for each prompt $x^n \in \mathcal{D}_b$
            \STATE Compute rewards $\{r_t^n\}_{t=1}^{|w^n|}$ for each sampled output $w^n$ by running $R_{\phi}$ \hfill
            \STATE Compute advantages $\{A_t\}_{t=1}^{|w^n|}$ and value targets $\{V^{\text{tar}}(s_t)\}_{t=1}^{|w^n|}$ for each $w^n$ with $V_\psi$
            \FOR{PPO iteration $= 1, \dots, \mu$}
                \STATE Update the policy model using PPO objective:
                \[
                \theta \gets \arg\max_\theta \frac{1}{|\mathcal{D}_b|} \sum_{n=1}^{|\mathcal{D}_b|} \frac{1}{|w^n|} \sum_{t=1}^{|w^n|} \min \left( \frac{P_\theta(a_t \mid s_t)}{P_{\theta_{\text{old}}}(a_t \mid s_t)} A_t, \, \text{clip}(v_t, 1 - \epsilon, 1 + \epsilon) A_t \right)
                \]
                \STATE Update the value model by minimizing a square-error objective:
                \[
                \psi \gets \arg\min_\psi \frac{1}{|\mathcal{D}_b|} \sum_{n=1}^{|\mathcal{D}_b|} \frac{1}{|w^n|} \sum_{t=1}^{|w^n|} \left( V_\psi(s_t) - V^{\text{tar}}(s_t) \right)^2
                \]
            \ENDFOR
        \ENDFOR
    \end{algorithmic}
\end{algorithm*}

\end{document}